\documentclass[english,twocolumn,amssymb,amsmath,superscriptaddress]{revtex4}
\usepackage[latin9]{inputenc}
\usepackage{graphicx,color}
\usepackage{hyperref}
\usepackage{babel}
\usepackage{booktabs}
\usepackage[normalem]{ulem}

\newcommand{\be}{\begin{equation}}
\newcommand{\ee}{\end{equation}}

\definecolor{mygreen}{rgb}{0,0.5,0}
\definecolor{myblue}{rgb}{0,0,0.75}
\definecolor{mymagenta}{cmyk}{0,1,0,0.12}

\begin{document}

\title{Dynamical Equilibration of Topological Properties}

\author{Andreas Kruckenhauser}
\affiliation{Department of Physics, University of Gothenburg, SE 412 96 Gothenburg, Sweden}
\affiliation{Institute for Theoretical Physics, University of Innsbruck, A-6020 Innsbruck, Austria}
\author{Jan Carl Budich}
\affiliation{Department of Physics, University of Gothenburg, SE 412 96 Gothenburg, Sweden}
\affiliation{Institute of Theoretical Physics, Technische Universit\"at Dresden, 01062 Dresden, Germany}	
\date{\today}

\begin{abstract}
We study the dynamical process of equilibration of topological properties in quantum many-body systems undergoing a parameter quench between two topologically inequivalent Hamiltonians. This scenario is motivated by recent experiments on ultracold atomic gases, where a trivial initial state is prepared before the Hamiltonian is ramped into a topological insulator phase. While the many-body wave function must stay topologically trivial in the coherent post-quench dynamics, here we show how the topological properties of the single particle density matrix dynamically change and equilibrate in the presence of interactions. In this process, the single particle density matrix goes through a characteristic level crossing as a function of time, which plays an analogous role to the gap closing of a Hamiltonian in an equilibrium topological quantum phase transition. As an exact case study exemplifying this mechanism, we numerically solve the quench dynamics of an interacting one-dimensional topological insulator.
\end{abstract}

\date{\today}

\maketitle

\section{Introduction }
\label{sec:int}
The recent experimental progress \cite{Aidelsburger2011,Sengstock,Ketterle2013,Aidelsburger2013,jotzu2014,Atala2014,Aidelsburger2015,Mancini:2015,Stuhl:2015} on synthetically realizing topological quantum matter \cite{HasanKane,XLReview} in ultracold atomic gases [see Ref. \cite{NathanReview} for a review] provides a natural platform to study the interplay between topological properties and non-equilibrium dynamics of quantum many-body systems. While experiments in conventional topological materials are typically aimed at probing equilibrium properties at low temperatures, the high degree of coherence in their atomic counterparts leads to inherently non-equilibrium protocols: Starting from a trivial initial state, the Hamiltonian is quenched into a topological regime, and physical properties are observed with a flexible and powerful toolbox of experimental techniques \cite{Bloch2008,Lewenstein2012} in the post-quench dynamics. A basic constraint in this scenario is, however, that the topology of the many-body wave function cannot change under coherent dynamics representing a local unitary transformation \cite{ChenGuWen2010}, thus leading to a trivial time-evolved state. Addressing this issue, it has recently been investigated \cite{foster2013quantum,Rigol2015, CooperQuenchCI, HamburgStateReconstruction, BudichHeyl2016, wang2016universal, YingDynamicalHall, RefaelDynamicalHall, WangNoneqChern, lezama2017one, SchuelerPhononChern, HamburgExperiment, HamburgLinking} how topological properties of the Hamiltonian still manifest in various experimental signatures, despite the trivial state.

\begin{figure}[!htb]
\begin{center}
\includegraphics[clip,width=0.45\columnwidth]{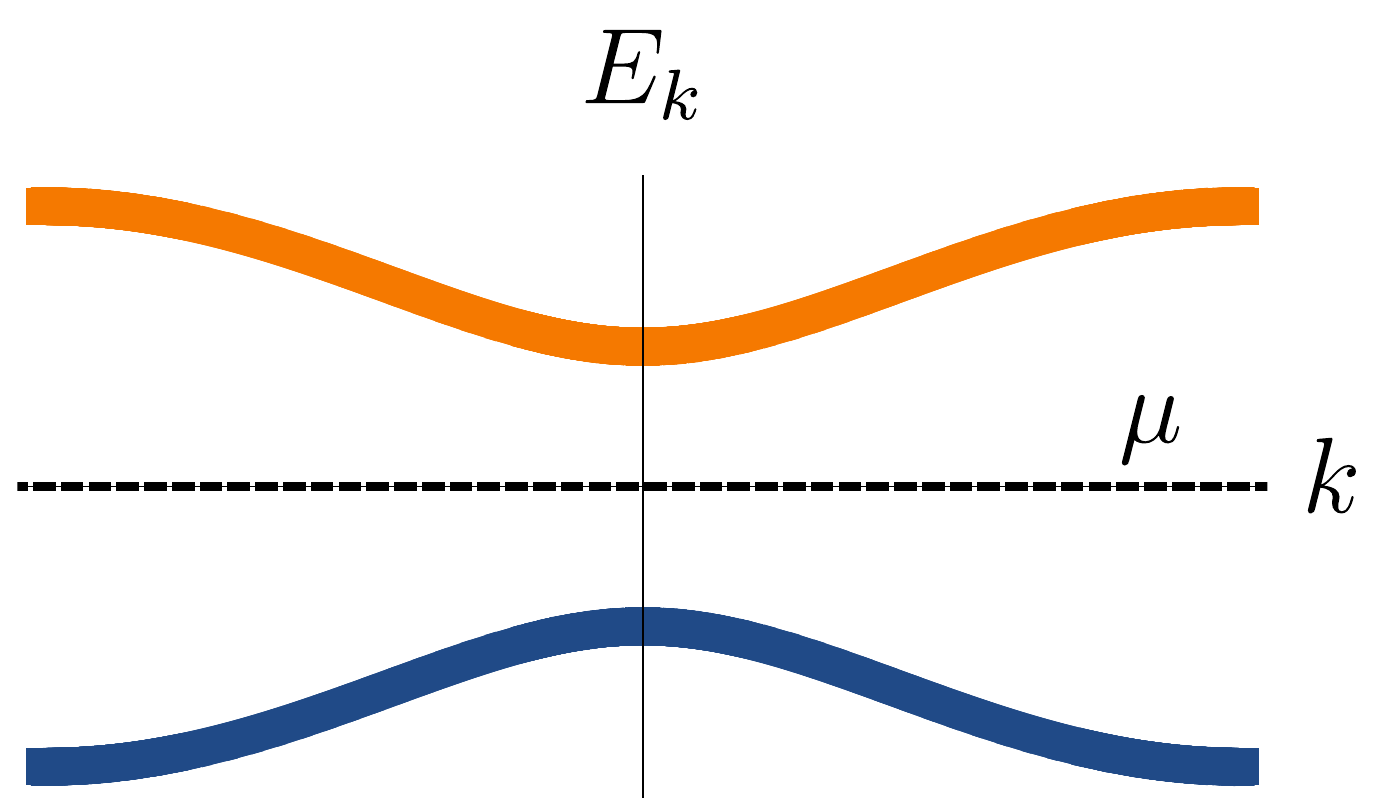}
\llap{\parbox[b]{7.5cm}{(a)\\\rule{0ex}{1.84cm}}}
\hspace{10pt}
\vspace{10pt}
\includegraphics[clip,width=0.45\columnwidth]{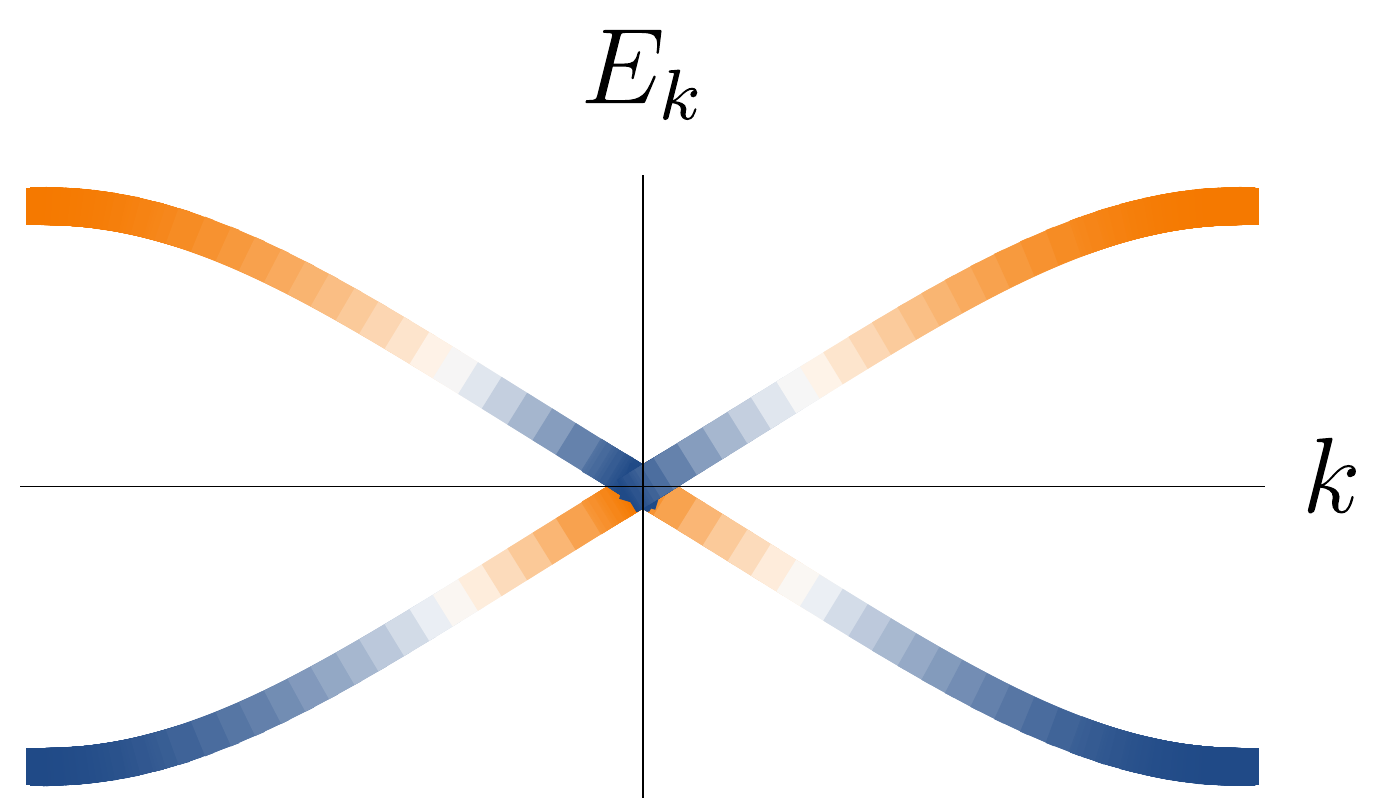}
\llap{\parbox[b]{7.5cm}{(b)\\\rule{0ex}{1.84cm}}}
\includegraphics[clip,width=0.45\columnwidth]{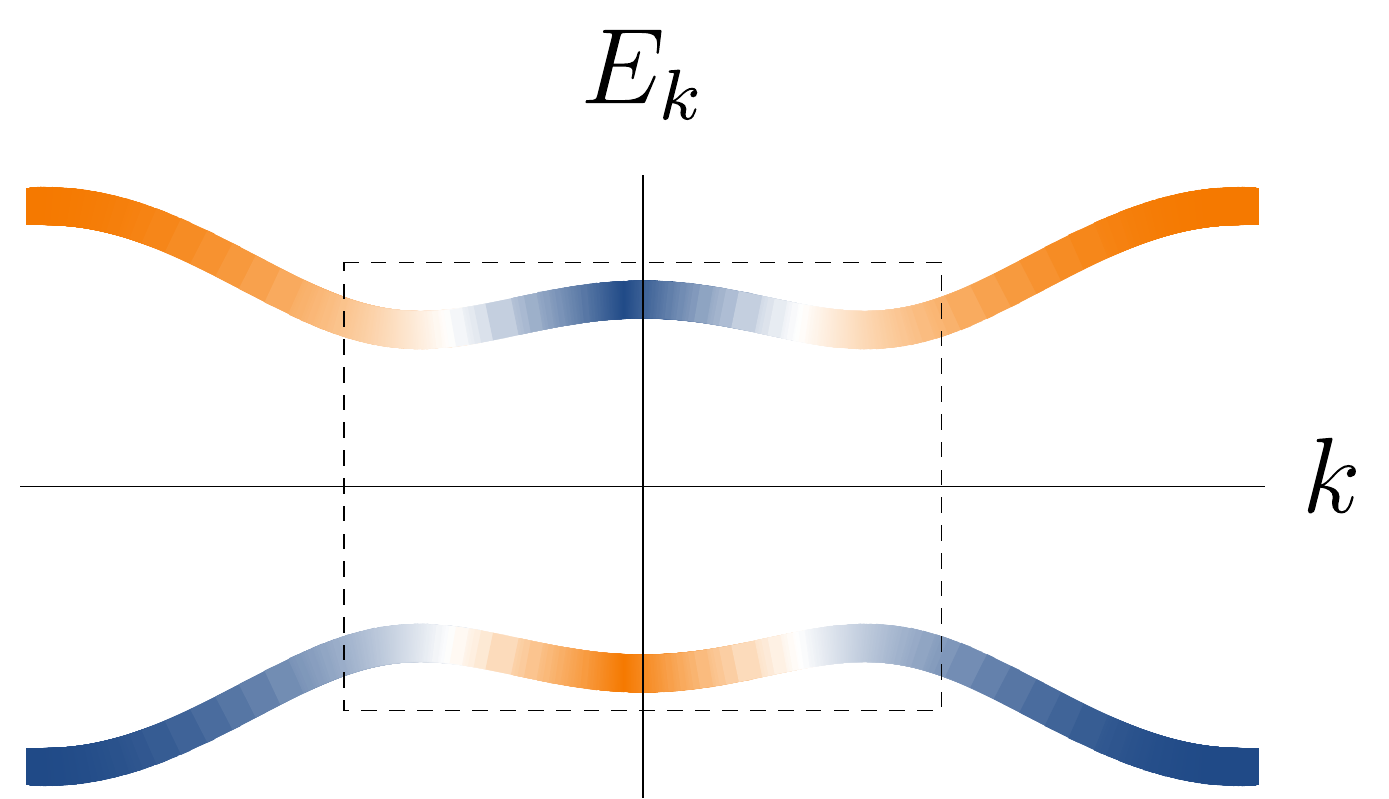}
\llap{\parbox[b]{7.5cm}{(c)\\\rule{0ex}{1.84cm}}}
\hspace{10pt}
\includegraphics[clip,width=0.45\columnwidth]{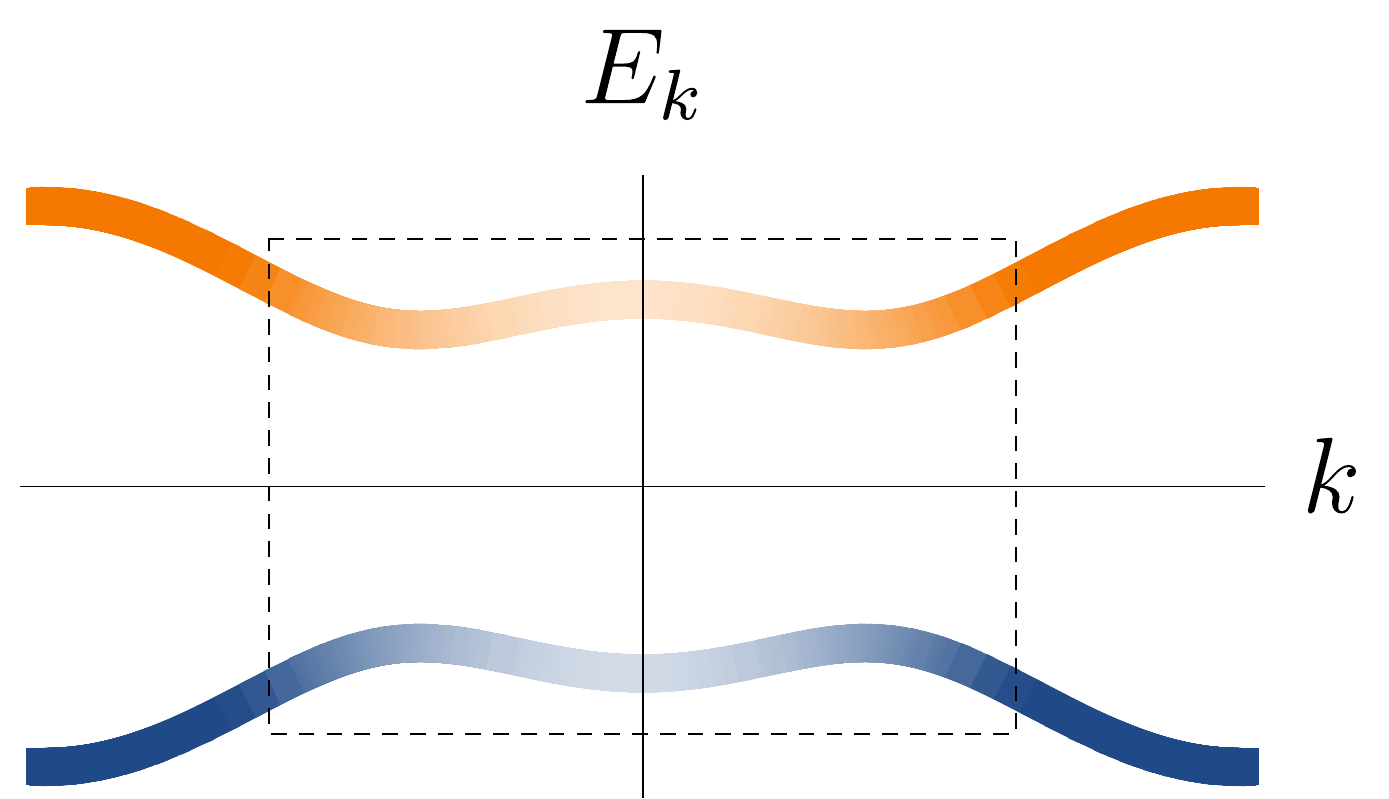}
\llap{\parbox[b]{7.5cm}{(d)\\\rule{0ex}{1.84cm}}}
\includegraphics[clip,trim={0 2cm 0 2cm},width=0.5\columnwidth]{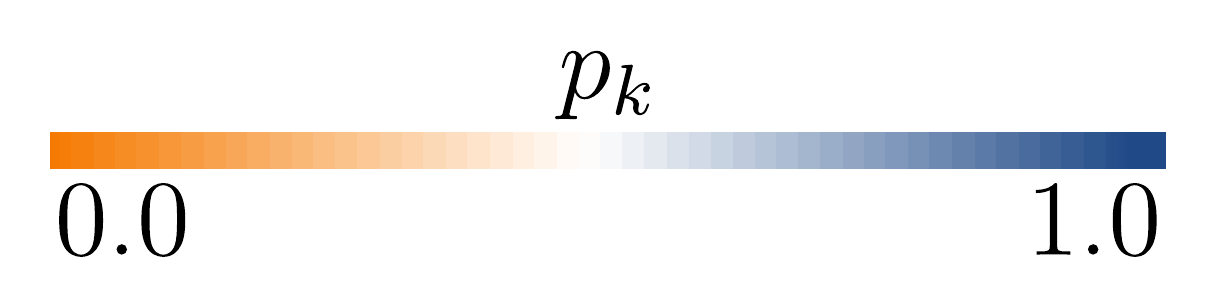}
\end{center}
\vspace{-10pt}
\caption{Thermalization of topological properties. The color code indicates the occupation probability $\text{p}_k$ of the Bloch states. (a) Insulating band structure in equilibrium at zero temperature, where the dashed line labeled by $\mu$ marks the chemical potential. (b) Gap closing at $k=0$ during a quench to a topological band structure. (c) Nonequilibrium situation with population inversion around $k=0$ immediately after the quench ($t=0$) to a topological (inverted) band structure. The dashed rectangular frame indicates the region in which significant excitations occur. (d) Thermal state of the topological band structure, dynamically equilibrated long time after the quench ($t\rightarrow \infty$).}
\label{fig:one}
\end{figure}

The purpose of this work is to investigate how {\emph{interaction-induced equilibration}}  can dynamically change topological properties of the single particle density matrix (SPDM) in coherent time evolution. In this scenario, even though the time evolution is Hamiltonian (coherent) at the many-body level, the dynamics of the SPDM, i.e. the reduced one-body state, effectively becomes dissipative due to interactions. This allows for a dynamical transition at a finite time after the quench, where the topological properties of the non-interacting part of the post-quench Hamiltonian and the single particle density matrix are reconciled. Hence, the SPDM exhibits close similarities to a topologically non-trivial thermal state in the long-time limit after the quench. By contrast, respecting the aforementioned constraint, the many-body wave function remains topologically trivial during the post-quench dynamics. This is because the coherent (symmetry-preserving) time-evolution with respect to a local Hamiltonian acts as a local unitary transformation which, by their very definition, cannot change (symmetry protected) topological properties of the wave-function \cite{ChenGuWen2010,footLUT}.\\

Below, we first present a general physical mechanism qualitatively explaining how this dynamical equilibration of topological properties of the SPDM can occur by redistribution of excitations in momentum space [see Fig.~\ref{fig:one}]. Then, we quantitatively exemplify and corroborate this qualitative picture with exact numerical studies on the non-equilibrium time-evolution of interacting one-dimensional topological insulators described by a Su, Schrieffer, and Heeger (SSH) model \cite{SSH,SSHReview} with two-body interactions. In particular, we explicitly demonstrate the aforementioned dynamical topological transition of the SPDM in real-time dynamics, and argue how the effective temperature of the equilibrated one-body state can be arbitrarily lowered by slowing down the quench. We emphasize that the topological discrepancy between the SPDM and the many-body state does not impose a finite lower bound on the resulting effective temperature. Furthermore, we note that this mechanism does not rely on finite size effects allowing to avoid a transition adiabatically, but instead even works in the thermodynamic limit.

The remainder of this article is structured as follows. In Sec. \ref{sec:qdi}, we generally analyze how interaction-induced equilibration can change topological properties of the SPDM. Sec. \ref{sec:dei} provides an exact numerical case study on the quench dynamics of interacting 1D topological insulators, and a concluding discussion is presented in Sec. \ref{sec:cd}.

\section{Quench-dynamics in interacting topological insulators.}
\label{sec:qdi}
We consider a fermionic lattice-periodic system with $n$ orbitals per site, initialized in the ordinary (non-topological) band-insulator ground state of an initial Hamiltonian $H_0^i$. Thereafter, the system undergoes a sudden or continuous parameter quench resulting in a topologically non-trivial post-quench Hamiltonian $H_0^f$. To fix a reference time, we assume the post-quench dynamics to start at $t=0$ \cite{foot1}. The time-dependent many-body state is denoted by $\lvert \psi(t)\rangle$. Taking into account interactions as described by the Hamiltonian $H_I^f$, $\lvert \psi(t)\rangle$ is given by ($\hbar=1$)
\begin{align}
\lvert \psi(t)\rangle = \text{e}^{-i(H_0^f+H_I^f)t}\lvert \psi(0)\rangle.
\label{eqn:psit}
\end{align}
While the many-body state $\lvert \psi(t)\rangle$ remains topologically trivial during the post-quench dynamics, the SPDM, defined as a function of the lattice momentum $k$ as 
\begin{align}
\rho_k^{\alpha\beta}(t)=\langle \psi(t)\rvert c_{k,\alpha}^\dag c_{k,\beta}\lvert \psi(t)\rangle,
\label{eqn:rhot}
\end{align}
where $c_{k,\alpha}$ annihilates a fermion at $k$ in orbital $\alpha=1,\ldots,n$, undergoes a dissipative dynamics, as the $c_{k,\alpha}$ operators obey non-linear field equations in the presence of interactions. This constitutes a crucial difference to the non-equilibrium dynamics in free fermionic systems, where all information about the many-body state, including in particular topological properties is contained in the SPDM. We note that the SPDM for a system initialized in a state with $n_o \le n$ occupied bands obeys the normalization condition $\text{Tr}\rho_k=n_o$, reflecting that $n_o$ single particle states are filled at every lattice momentum. The name SPDM is motivated (even for $n_o>1$) from the fact that the SPDM is formed of matrix elements of a one-body operator [see Eq. (\ref{eqn:rhot})]. Here, we are interested in the intriguing dynamical process of a change in the topological properties \cite{NathanReview,DiehlTopDiss,BardynTopDiss,DissCI,TopDens} of the SPDM during the real-time evolution after the quench. An intuitive way of defining the topological properties of the SPDM is to formally interpret $\rho_k(t)$ as a thermal state (with zero chemical potential and unit temperature) of a (fictitious) free Hamiltonian $\tilde H_k(t)$, defined via $\rho_k(t)=\text{e}^{-\tilde H_k(t)}/\tilde Z_k(t)$ with $\tilde Z_k(t)=\frac{1}{n_o}\text{Tr}\left\{\text{e}^{-\tilde H_k(t)}\right\}$. The topological properties of $\rho_k(t)$ are then simply inherited from $\tilde H_k(t)$ and are well defined as long as $\tilde H_k(t)$ is gapped. We emphasize that $\tilde H_k(t)$ should not be mixed up with the actual instantaneous Hamiltonian of the system at time $t$, since the system is far from equilibrium before approaching its thermalized steady state. 

While at a qualitative level the mechanism that leads to the dynamical equilibration of the topological properties of the SPDM is quite generic, we illustrate it for concreteness with a two-band model, i. e. $\alpha=1,2$ in Eq. (\ref{eqn:rhot}), in one spatial dimension (1D) [see Fig.~\ref{fig:one}], which is also be the relevant case for our numerical study below. Right after the quench at $t=0$, the SPDM is topologically trivial, but exhibits a population inversion around the critical momentum $k_c$ at which the gap closing occurred during the quench due to the topological transition in the Hamiltonian [see Fig.~\ref{fig:one}(c)]. This population inversion is essential for the un-correlated trivial state at $t=0$ to compensate for the band-inversion of the now topologically non-trivial Hamiltonian. However, the accumulation of excitations around $k_c$ is not stable under two-body scattering, and is thus dynamically redistributed (equilibrated) according to the energy band dispersion of $H_0^f$ by weak interactions. As a consequence, the SPDM $\rho_k(t)$  approaches a thermal state of the topologically non-trivial $H_0^f$ [see Fig.~\ref{fig:one}(d)], thus undergoing a topological transition at some $t_c>0$. This transition happens via a level crossing in the SPDM, also referred to as a purity gap closing \cite{NathanReview,DiehlTopDiss,BardynTopDiss,DissCI,TopDens}, which can be seen as the analog of a band inversion for density matrices. This analogy becomes intuitively clear in terms of the aforementioned (fictitious) free Hamiltonian $\tilde H_k(t)$: At the critical time $t_c$, the level crossing in $\rho_k(t)$ then simply corresponds to a topological transition with an energy gap closing of $\tilde H_k(t)$.

As the density of excitations created during the quench is non-universal, the effective temperature of the topologically non-trivial thermal SPDM long time after the quench can be tuned with the quench parameters. In particular, slowing down the quench systematically reduces the energy pumped into the system due to simple Landau-Zener physics: Excitations are only created in a window around the critical momentum $k_c$ where the system cannot adiabatically stay in the ground state.  Independent of system size, this window shrinks with a slower ramp velocity and so does the total density of excitations and the effective finite temperature. Still, since we are concerned with the thermodynamic limit, excitations are created with probability one right at $k_c$, leading to the mentioned population inversion no matter how slow the quench. Hence, without the interaction induced equilibration discussed here, the SPDM would remain topologically trivial at all times.

\section{Dynamical equilibration in interacting SSH model. } 
\label{sec:dei}
We now exemplify the dynamical equilibration of topological properties with a concrete case study on quenches in an interacting SSH model, i.e. a 1D topological insulator on a lattice with unit lattice constant protected by a particle hole constraint (PHC). The non-interacting part of the model Hamiltonian reads as
\begin{align}
H_0(t) =& -\sum_j \left [J(t) b_j^\dag a_j + \frac{d+\tau}{2}b_j^\dag a_{j+1} + \frac{d-\tau}{2}a_j^\dag b_{j+1}\right]\nonumber\\
&+\text{h.c.},
\label{eqn:freessh}
\end{align}

\begin{figure}[htp!]
\begin{centering}
\includegraphics[clip,trim=2.2cm 0.2cm 01cm 0.2cm,width=0.97\columnwidth]{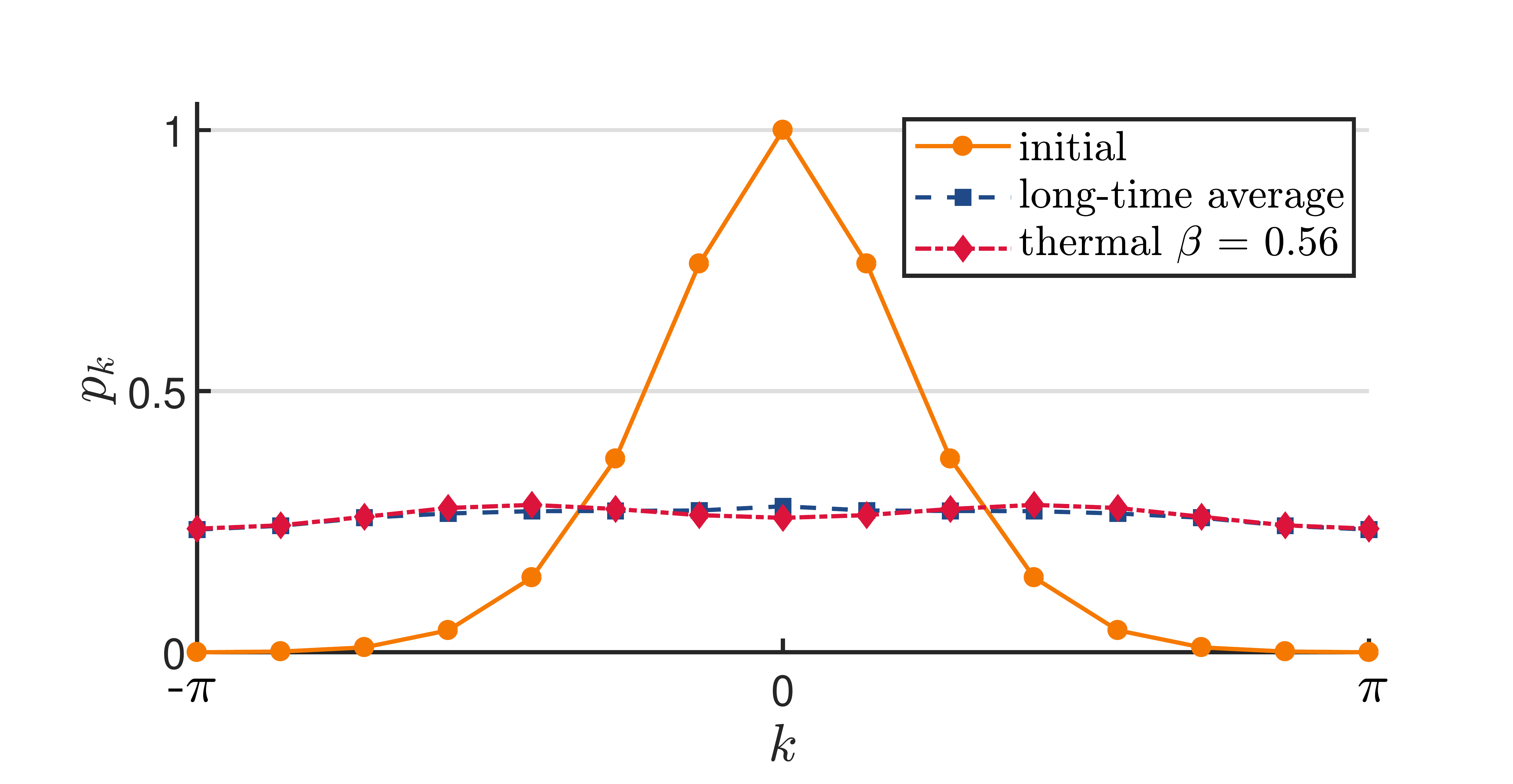}
\llap{\parbox[b]{16.6cm}{(a)\\\rule{0ex}{3.65cm}}}
\includegraphics[clip,trim=2.2cm 0.2cm 01cm 0.2cm,width=0.97\columnwidth]{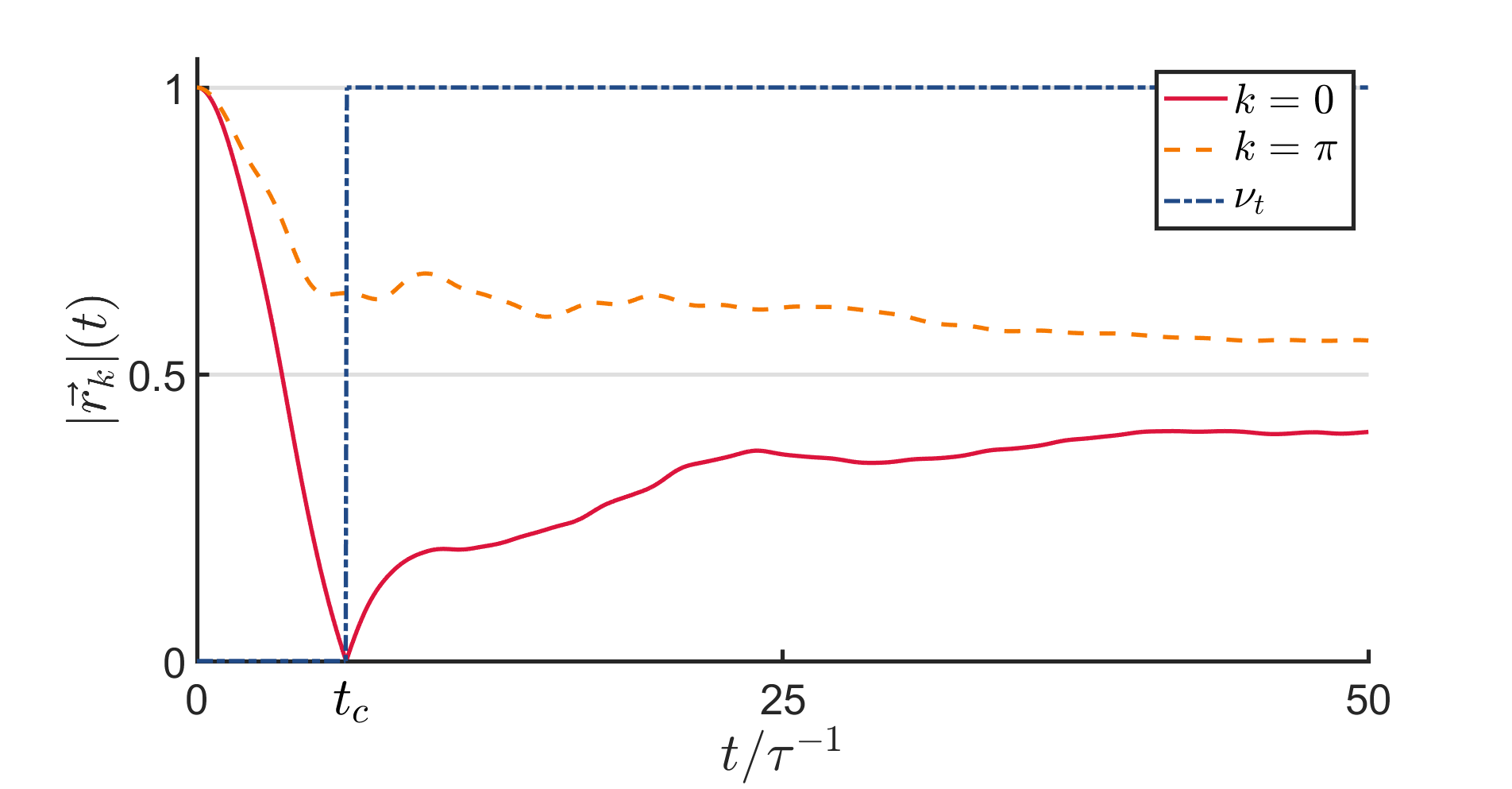}
\llap{\parbox[b]{16.6cm}{(b)\\\rule{0ex}{3.89cm}}}
\includegraphics[clip,trim=2.2cm 0.2cm 01cm 0.2cm,width=0.97\columnwidth]{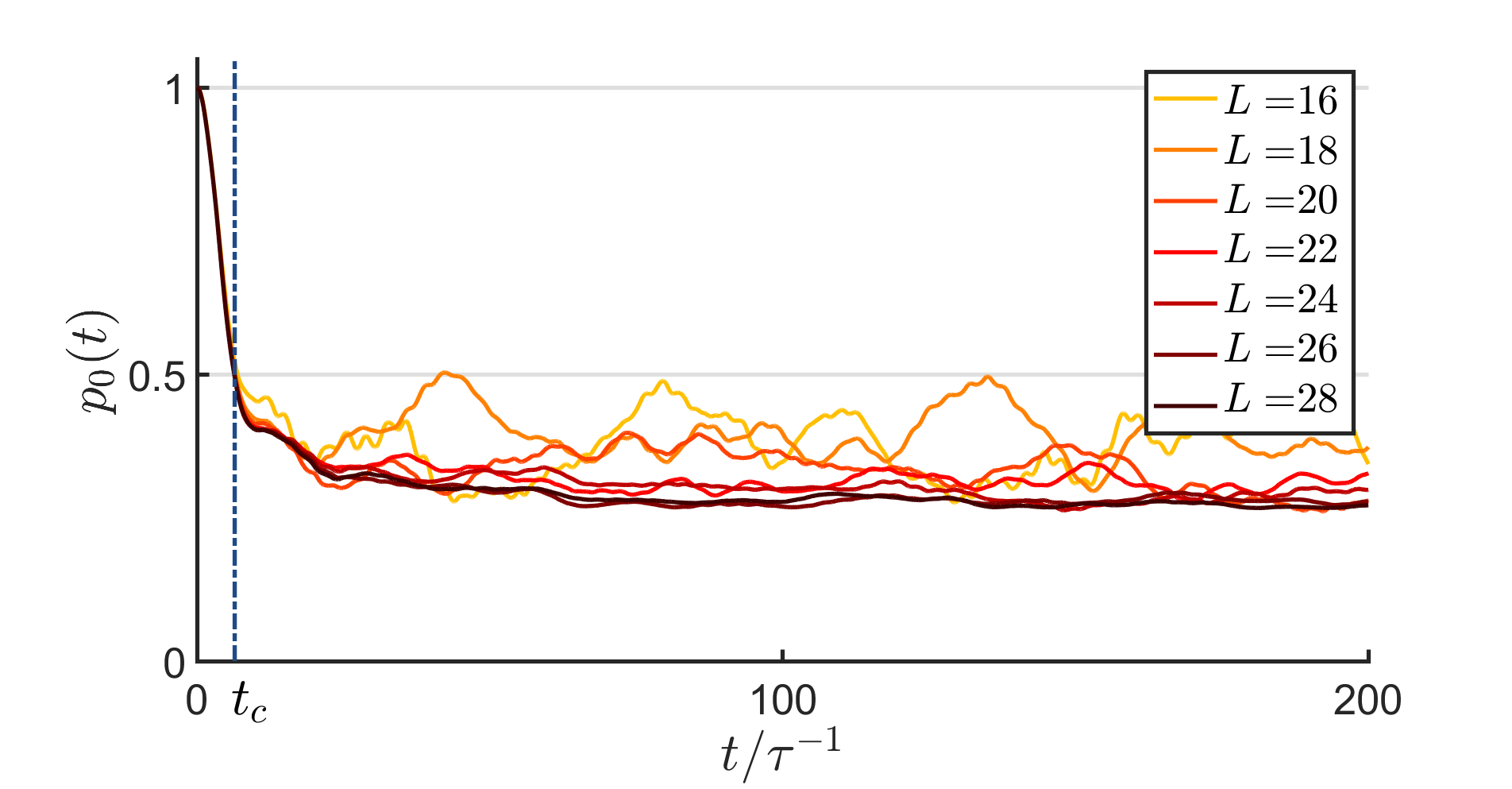}
\llap{\parbox[b]{16.6cm}{(c)\\\rule{0ex}{3.89cm}}}
\caption{\label{fig:three} Numerical results. (a) Comparison of the probability $p_k$ of excitations to the upper Bloch band as a function of momentum $k$ at $t=0$ (orange solid), long time after the quench (blue dashed), and for a thermal state of $H_i^f$ with the same energy expectation value (red dashed-dotted) for a system with $L=28$ spinless sites. (b) Purity $\lvert\vec r_{k=0}\rvert(t)$ of the SPDM [see Eq. (\ref{eqn:rhotwoband})] at $k=0$ (red solid) and $k=\pi$ (orange dashed) as a function of time. The transition time is marked with $t_c$ and the time-dependent value of the topological index $(-1)^{\nu_t}=\text{sgn}(r_0^x(t)r_\pi^x(t))$ is indicated by the blue dashed-dotted line. (c) Finite size dependence of the time-dependent probability $p_{k=0}$ of excitations to the upper Bloch band for system sizes between $L=16$ and $L=28$ sites. Parameters in all plots are $d=0.84\tau$, $V=0.30\tau$, $J(t\rightarrow -\infty)=-1.95\tau$, $J(t=0)=-0.05\tau$ and the quench is parameterized as  
$J(t)=(J(0)-J(-\infty))\left(\text{tanh}(1.5/\tau \cdot t)+1\right)/2+J(-\infty)$.}
\end{centering}
\vspace{-0pt}
\end{figure}

where $a_j$ ($b_j$) annihilates a spinless fermion on sublattice A (B) of the super-site $j$, and $J(t),d,\tau$ are real hopping parameters. The quench is encoded in the time-dependence of $J(t)$, where $J(t\rightarrow -\infty)$ characterizes the initial Hamiltonian $H_0^i$, whereas $J(t=0)$ enters the final Hamiltonian $H_0^f$. In momentum space, $H_0$ may be expressed as
\begin{align}
H_0(t)= \sum_k c_k^\dag \,\left[\vec h_k(t) \cdot \vec \sigma \right]\,c_k,
\end{align}  
where $c_k=(c_{k,a},c_{k,b})$ denotes the spinor in A-B sublattice space of annihilation operators at lattice momentum $k$, the vector $\vec \sigma$ denotes the standard Pauli-matrices acting in A-B sublattice space, and $\vec h_k(t)=\left(-J(t)-\tau \cos(k),-d\sin(k),0\right)$. The PHC of this model is tantamount to the anti-commutation relation $\left\{H_0,\sigma_z K\right\}=0$, where $K$ denotes complex conjugation. The (equilibrium) topological invariant \cite{Kitaev2001} can be readily evaluated by looking at the so called real lattice momenta $k=0,\pi$ which are characterized by $k=-k$. There, the PHC boils down to $h_k^y=h_k^z=0$, and the topological invariant is simply given by $(-1)^\nu = \text{sgn}(h_0^xh_\pi^x)$, where $\nu=1$ represents the topologically non-trivial phase and $\nu = 0$ characterizes a trivial state. $\nu$ can be understood as ($1/\pi$ times) the Zak-Berry phase \cite{ZakPol} which measures the charge polarization relative to the position of the lattice sites, and which is quantized due to the PHC \cite{HatsugaiQuantizedBerry}. From $h_k^x=-J-\tau\cos(k)$, we immediately read off that the topological regime is determined by $\lvert \tau\rvert >\lvert J\rvert$. Hence, when quenching into a topological regime, we initialize the system in the ground state of $H_0^i$ with $\lvert J(-\infty)\rvert>\lvert\tau\rvert>0$ and quench to a final set of parameters for $H_0^f$ satisfying $\lvert\tau\rvert>\lvert J(0)\rvert>0$. The two-body interactions during the post-quench dynamics are described by the Hamiltonian 
\begin{align}
H_I^f = V\sum_j(n^a_j\,n^b_{j+1}+n^a_{j+1}\,n^b_j), 
\label{eqn:intssh}
\end{align}
where $n^a_j=a_j^\dag a_j, n^b_j=b_j^\dag b_j$, and the time-dependence of the many-body state $\lvert\psi(t)\rangle$ is given by Eq. (\ref{eqn:psit}). Starting from all possible interaction terms up to nearest neighbor range, we note that the intra-sublattice interaction $W_1\sum_j n^a_j\, n^a_{j+1}$ and $W_2 \sum_jn^b_j\, n^b_{j+1}$ breaks the PHC that protects the topological invariant $\nu$, unless $W_1=W_2$. However, as both the on-site interaction $U\sum n^a_j \,n^b_j$ and the $W_1=W_2$ term are found to be much less efficient (numerically manifesting in extremely long equilibration time-scales) for the thermalization process than $V$, we focus on the case $U=W_1=W_2=0\ne V$, as described by Eq. (\ref{eqn:intssh}). Using the Krylov time-propagation method \cite{Nauts1983}, we numerically compute $\lvert \psi(t)\rangle$ exactly for systems at half-filling with periodic boundary conditions, consisting of up to $L=28$ spinless sites, i.e. up to $14$ super-cells with $14$ particles.

To evaluate the (time-dependent) topological properties of the SPDM, we re-express $\rho_k(t)$ [see Eq. (\ref{eqn:rhot})] as
\begin{align}
\rho_k(t)=\frac{1}{2}\left(1-\vec r_k(t)\cdot \vec\sigma\right).
\label{eqn:rhotwoband}
\end{align}  
In this notation, the initial one-body state, which is simply the pure ground state of $H_0^i$  is characterized by $\vec r_k(-\infty)=\hat h_k(-\infty)$, where $\hat h_k(-\infty)=\vec h_k(-\infty)/\lvert \vec h_k(-\infty)\rvert$.  As long as $r_k\equiv \lvert \vec r_k\rvert>0$, we can formally express the SPDM as a thermal state of a \emph{gapped} Hamiltonian $\tilde H_k(t)$ obeying the same PHC, and the topological invariant of $\rho_k(t)$ is simply given by $(-1)^{\nu_t}=\text{sgn}(r_0^x(t)r_\pi^x(t))$. The invariant $\nu_t$ is found to concur with the ensemble geometric phase of the SPDM as introduced in Ref. \cite{EnsembleGP}.  As mentioned before, for the SPDM to undergo a dynamical topological transition, i.e. for $\nu_t$ to change sign as a function of time, a so called purity gap closing needs to occur at a critical time $t_c$. There, $r_k(t_c)=0$ for $k=0$ or $k=\pi$ which is equivalent to the fictitious $\tilde H_k(t_c)$ exhibiting a gap closing.

In Fig.~\ref{fig:three}, we show numerical data exemplifying such a process of dynamical equilibration of topological properties. Panel (a) compares the momentum distribution functions of the SPDM at different times to that of a thermal state with respect to $H_0^f$. The non-equilibrium distribution of the topologically trivial SPDM immediately after the quench (orange solid) exhibits the aforementioned characteristic population inversion around $k=0$. The time-averaged SPDM long time after the quench (blue dashed) shows good agreement with the thermal state (red dotted) of $H_0^f$ that is matched in temperature to have the same energy expectation value. Panel (b) shows the time-dependent purity $\lvert\vec r_{k=0}\rvert(t)$ [see Eq. (\ref{eqn:rhotwoband})] at zero momentum, exhibiting the aforementioned purity-gap closing at a critical time $t_c$, where topological index $\nu_t$ of the SPDM becomes non-trivial. Panel (c) displays the influence of finite size effects on the equilibration of the SPDM which are most prominent at momentum $k=0$. The persistent fluctuations in the occupation number observed for small systems are strongly suppressed with increasing system size. This indicates that the largest simulated system captures the thermodynamic limit quite well for the chosen quench parameters.

\section{Concluding Discussion.}
\label{sec:cd}
 Our analysis shows how low-temperature topological insulator states  of the SPDM can be generically prepared in a non-equilibrium fashion by slowly quenching a weakly interacting system. Since the momentum distribution functions contained in the SPDM are directly accessible by standard time-of flight measurements and determine numerous physical properties including important response functions, our findings are of immediate experimental relevance. On a more general note, the discussed quench protocol gives a partial (note that many-body state is bound to remain trivial) solution to the general issue of state preparation in synthetic materials, one of the key challenges in the field of quantum simulation. Complementary approaches towards the dynamical preparation of topological states include the notion of dissipative state preparation \cite{NathanReview,DiehlDissPrep,Verstraete2008,Weimer2010,barreiroNature,Krauter2011,kapit14} in open quantum systems with so called engineered dissipation, and adiabatic passage in small systems \cite{CiracAdiabatic,adiabaticTO,PollmannAdiabatic} exhibiting a sufficient finite size gap. The scenario discussed in the present work, by contrast, does not rely on finite size effects and uses the interaction-induced intrinsic dissipation seen by one-body observables in a closed many-body system that acts as its own bath. The final effective temperature of the SPDM long time after the quench is only determined by the quench parameters. Thus, arbitrarily low temperatures can in principle be achieved by slowing down the parameter ramp. However, in real experiments the finite lifetime of synthetic material systems such as ultracold atoms in optical lattices will put a lower bound to the accessible temperature.

We have focused on bulk topological properties in translation invariant (periodic) systems. For systems with open boundaries, it is natural to ask if and how the sub-gap edge states of topological insulators equilibrate dynamically. In the present 1D model (see Eq. (\ref{eqn:freessh})), the zero-energy edge states are energetically isolated from the bulk, and are hence found to not thermalize  efficiently at all. In higher spatial dimensions than 1D, the spectrum of the metallic edge states continuously merges into the bulk, which may give rise to their more efficient interaction-induced equilibration. However, a fully microscopic study of such interacting higher dimensional systems is beyond the scope of our present exact diagonalization analysis.   

The thermalization processes discussed in this work pertain to topological properties of the SPDM, i.e. the one-body state, while the many-body wave function remains topologically trivial during the coherent post-quench dynamics. For topological phases such as topological insulators and superconductors that have representatives within the class of free fermionic models, the preparation of an effective low temperature SPDM in weakly interacting systems may already be an important step towards observing low temperature physics in such systems. However, at least for inherently strongly correlated topological phases such as fractional quantum Hall systems, where crucial properties such as fractionalized excitations cannot be fully understood at the level of single particle Green's functions, genuine non-equilibrium phenomena beyond the scope of this work are expected. Analyzing the interplay of topology and quench-dynamics in such systems is an interesting subject of future research. 

\emph{Acknowledgements. } We acknowledge discussions with  Y. Hu, M. Heyl, and L. Pastori. The numerical calculations were performed on resources at the Chalmers
Centre for Computational Science and Engineering (C3SE) provided by the Swedish National Infrastructure for Computing (SNIC).

\bibliographystyle{apsrev}

\begin{thebibliography}{10}
\bibitem{foot1}
{For a sudden quench $t=0$ is the instant in time where the parameter change happens, for a continuous ramp, $t=0$ hallmarks the time where the parameter ramp is just concluded.}
\end{thebibliography}

\end{document}